\documentclass{ws-ijmpcs}
\newcommand{\ben}{\begin{eqnarray}}
\newcommand{\een}{\end{eqnarray}}
\newcommand{\nnu}{\nonumber\\}
\newcommand{\bef}{\begin{figure}[htb]\centering}
\newcommand{\eef}{\end{figure}}

\begin{document}

\markboth{Zhong-Bo Kang and Jian-Wei Qiu}
{Scale dependence of twist-3 correlation functions}

\catchline{}{}{}{}{}
%
%%%%%%%%%%%%%%%%%%%%%%%%%%%%%%%%%%%%%%%%%%%%%%%%%%%%%%%%%%%%%%%%%%%%

\title{QCD evolution of naive-time-reversal-odd quark-gluon correlation functions\footnote{Presented by J.-W. Qiu.}
}

\author{Zhong-Bo Kang$^{1}$ and Jian-Wei Qiu$^{2,3}$ }

\address{$^1$Theoretical Division, Los Alamos National Laboratory\\
                 Los Alamos, NM 87545, USA \\
                 $^2$Physics Department,
                 Brookhaven National Laboratory,
                 Upton, NY 11973, USA \\
                 $^3$C.N. Yang Institute for Theoretical Physics,
                 Stony Brook University\\
                 Stony Brook, NY 11794, USA
}

\maketitle

\begin{history}
\received{Day Month Year}
\revised{Day Month Year}
\end{history}

\begin{abstract}
In this talk, we examine the existing calculations of QCD evolution kernels for the scale dependence of two sets of twist-3 quark-gluon correlation functions, $T_{q,F}(x,x)$ and $T^{(\sigma)}_{q,F}(x,x)$, which are the first transverse-momentum-moment of the naive-time-reversal-odd Sivers and Boer-Mulders function, respectively.  The evolution kernels at the leading order in strong coupling constant $\alpha_s$ were derived by several groups with apparent differences.  We identify the sources of discrepancies and are able to reconcile the results from various groups.

\end{abstract}

\ccode{PACS numbers: 11.10.Hi, 12.38.Bx,12,39.St, 13.88.+e}

%%%%%%%%%%%%%%%%%%%%%%%%%%%%%%%%%
\section{Introduction}	

Quantum Chromodynamics (QCD), as a fundamental theory of strong interaction dynamics, has been very successful in both lattice QCD calculations of static hadron properties and perturbative calculations of short-distance dynamics observed in high energy scattering processes.  In particular, with the aid of QCD factorization theorem\cite{CSS-fac}, QCD perturbation theory has done an excellent job in interpreting data from high energy scattering with identified hadrons, whose nonperturbative dynamics at the hadronic scale ($\sim$~1/fm) are represented by process independent and well-determined parton distribution and fragmentation functions\cite{CTEQPDF}\cdash\cite{MRSTPDF}.  The predictive power of QCD perturbation theory for cross sections with identified hadrons is an immediate consequence of QCD factorization and our abilities to predict the variation of these nonperturbative but universal distribution and fragmentation functions when they are probed at different hard momentum scales.  The success of QCD perturbation theory provides the well-controlled and calibrated sub-femtometer or even attometer ``scope"  to probe partonic structure and dynamics inside a hadron of bounded quarks and gluons.

On the other hand, the large size of observed transverse single-spin asymmetries (SSAs),
$A_N \equiv (\sigma(s_T)-\sigma(-s_T))/(\sigma(s_T)+\sigma(-s_T))$, 
defined as the ratio of the difference and the sum of the cross sections when the spin of one of the identified hadron $s_T$ is flipped, came as a surprise, and had posed a challenge for researchers in this field for some time\cite{Kane:1978nd}.
From the parity and time-reversal invariance of  the strong interaction dynamics, the measured large asymmetries in high energy collisions should be directly connected to the transverse motion of partons inside a polarized hadron. Transverse spin physics has attracted tremendous attention from both experimental and theoretical communities in recent years\cite{D'Alesio:2007jt}. 
Experimental measurements of the asymmetries and the investigation to understand the underlying dynamics have provided and will continue to provide us new opportunities to explore QCD and the hadron structure far beyond what we have been able to achieve.

Two complementary QCD-based approaches have been proposed to analyze the physics behind the measured asymmetries: the transverse momentum dependent (TMD) factorization approach\cite{TMD-fac}\cdash\cite{boermulders} and the collinear factorization approach\cite{Efremov,qiu,koike}. In the TMD factorization approach, the asymmetry was attributed to the spin and transverse momentum correlation between the identified hadron and the active parton, which are represented by the TMD parton distribution or fragmentation function.  On the other hand, in the collinear factorization approach, all active partons' transverse momenta are integrated into the collinear distributions, and the explicit spin-transverse momentum correlation in the TMD approach is now included into the high twist collinear parton distributions or fragmentation functions.  The asymmetry in the collinear factorization approach is represented by twist-3 collinear parton distributions or fragmentation functions, which have no probability interpretation, and could be interpreted as the quantum interference between a collinear active quark (or gluon) state in the scattering amplitude and a collinear quark (gluon)-gluon composite state in its complex conjugate amplitude.  The relevant TMDs and the twist-3 quark-gluon correlation functions, while they are both nonperturbative, are closely related to each other.  In general, the collinear twist-3 correlation functions are proportional to the parton transverse momentum $k_\perp$-moments of TMDs up to the uncertainty of ultraviolet renormalization of composite operators defining the moments of TMDs.  The TMD factorization approach is more suitable for evaluating the SSAs of scattering processes with two very different momentum transfers, $Q_1\gg Q_2 \gtrsim \Lambda_{\rm QCD}$, where the $Q_2$ is sensitive to the active parton's transverse momentum, while the collinear factorization approach is more relevant to the SSAs of scattering cross sections with all observed momentum transfers hard and comparable: $Q_i\sim Q\gg \Lambda_{\rm QCD}$. Although the two approaches each have their own kinematic domain of validity, they are consistent with each other in the regime where they both apply\cite{UnifySSA,Bacchetta:2008xw}.  

Both factorization approaches necessarily introduce a factorization scale, $\mu\gg\Lambda_{\rm QCD}$, to separate the calculable short-distance perturbative dynamics from the long-distance nonperturbative physics of the observed cross sections or the asymmetries.  Since the physical observables, the cross sections or the asymmetries, are independent of the choice of the factorization scale, the scale dependence of the nonperturbative distributions\cite{Aybat:2011ge}\cdash\cite{Kang:2010xv}, either TMD distributions or twist-3 collinear distributions, must match the scale dependence of corresponding  perturbative hard parts.  That is, the factorization scale dependence of the nonperturbative distributions is perturbatively calculable and is a prediction of QCD perturbation theory when $\mu\gg\Lambda_{\rm QCD}$. For example, the scale dependence of the leading power parton distributions obeys DGLAP evolution equations whose evolution kernels are perturbatively calculable, and has been very successfully tested when the scale varies from a few GeV to the hundreds of GeV.  

In this talk, we present a general evolution equation of the twist-3 quark-gluon correlation functions that are responsible for the SSAs in the collinear factorization approach\cite{Kang:2008ey}, and a detailed discusion on the evolution of two twist-3 quark-gluon correlation functions $T_{q,F}(x, x)$ and $T_{q,F}^{(\sigma)}(x, x)$.  These two correlation functions are defined as \cite{qiu,Kang:2012em}
\ben
T_{q, F}(x, x)&=&\int\frac{dy_1^- dy_2^-}{4\pi}e^{ixP^+y_1^-}
\langle P,s_T|\bar{\psi}_q(0)\gamma^+\left[ \epsilon^{s_T\alpha n\bar{n}}F_\alpha^{~ +}(y_2^-)\right] \psi_q(y_1^-)|P,s_T\rangle\, ,
\label{Tq}
\\
T_{q, F}^{(\sigma)}(x, x)&=&\int\frac{dy_1^- dy_2^-}{4\pi}e^{ixP^+y_1^-}
\frac{1}{2}\sum_{s_T} \langle P,s_T|\bar{\psi}_q(0)\left[ \sigma^{\alpha +} F_\alpha^{~ +}(y_2^-)\right] \psi_q(y_1^-)|P,s_T\rangle \, ,
\label{Tqs}
\een
where the gauge links between field operators are suppressed and $\epsilon^{0123}=1$ is used.  These two correlation functions
are equal to the first $k_\perp$-moment of the two well-known naive-time-reversal-odd TMDs, the Sivers function $f_{1T}^\perp(x, k_\perp^2)$\cite{Siv90} and the Boer-Mulders function $h_1^{\perp}(x, k_\perp^2)$\cite{boermulders}, respectively.  The scale dependence of these two twist-3 quark-gluon correlation functions has been studied recently by several groups\cite{Kang:2008ey,Zhou:2008mz,Vogelsang:2009pj,Braun:2009mi}. However, there are discrepancies between these results, particularly for the evolution of  $T_{q, F}(x, x)$ (also often refer to as Efremov-Teryaev-Qiu-Sterman (ETQS) function).  We show in this talk\cite{Kang:2012em} that these discrepancies can be resolved, and in addition we also present the evolution equations for the quark-gluon correlation function $T_{q, F}^{(\sigma)}(x, x)$.  

%%%%%%%%%%%%%%%%%%%%%%%%%%%%%%%%%
\section{Collinear factorization approach to SSAs}

In this section, we give a general discussion of collinear factorization approach to perturbative QCD treatment of SSAs of cross sections with one large momentum transfer $Q$.

\begin{figure}[h]
\begin{center}
\includegraphics[width=11cm]{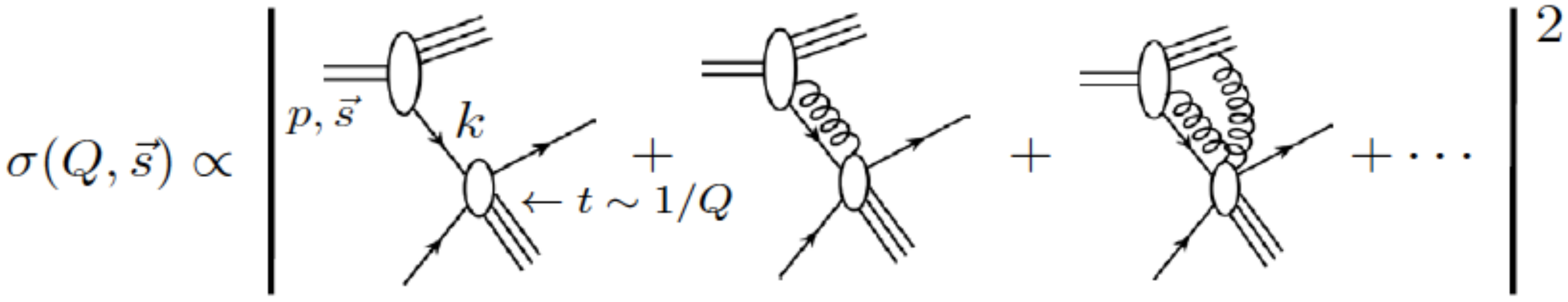}
\end{center}
%\vspace*{8pt}
\caption{A schematic illustration of the source of contribution to the scattering cross section between a simple particle and a hadron of momentum ${p}$ and spin $\vec{s}$. 
\label{fig:xsec}}
\end{figure}

A scattering cross section on a hadron is proportional to the square of the hadron's scattering amplitude, which sums over all partonic scattering amplitudes that share the same initial- and final-states but with different number of active partons, as illustrated in Fig.~\ref{fig:xsec}.  With one large momentum transfer $Q$, the hard scattering is localized to a distance scale of $1/Q$.   Since pulling an extra physically polarized parton into the localized collision point is suppressed by the power of $1/Q$, the cross section for a hadron $A$ to scatter off a hadron $B$ with a large momentum transfer $Q$  can be expanded in a power series in $1/Q$,
\begin{eqnarray}
\sigma_{AB}(Q,\vec{s}) 
&=&
\sigma_{AB}^{\rm LP}(Q,\vec{s}) + \frac{Q_s}{Q}\, \sigma_{AB}^{\rm NLP}(Q,\vec{s})+ ...
\label{eq:pexp}\\
&\approx &
H_{ab}^{\rm LP}\otimes f_{a/A} \otimes f_{b/B} 
\label{eq:fac-lp} \\
&\ &
+ \frac{Q_s}{Q}\, \left[
H_{ab}^{{\rm NLP-}A}\otimes T_{(a,F)/A}(\vec{s})  \otimes f_{b/B}  \right.
\label{eq:fac-Tf} \\
&\ &
\hskip 0.3in \left.
+ H_{ab}^{{\rm NLP-}B}\otimes h_{a/A}(\vec{s}) \otimes F_{(b,F)/B} ^{(\sigma)}
+ ... \right] + ...
\label{eq:fac-Tfs}
\end{eqnarray}
where $Q_s^2 \sim \langle k^2\rangle,  \langle k_T^2\rangle$ represents a characteristic scale of the power corrections.  According to the QCD collinear factorization theorem\cite{CSS-fac}, the leading power contribution to the cross section of the hadron-hadron collisions in Eq.~(\ref{eq:pexp}) is given by the square of the scattering amplitude with one active {\em collinear} parton from each colliding hadron (plus any number of collinear and longitudinally polarized gluons responsible for the gauge links), like the first term of the scattering amplitude in Fig.~\ref{fig:xsec}.  The contribution can be factorized into a convolution of a localized and perturbatively calculable hard part $H^{\rm LP}_{ab}$ from the collision between partons $a$ and $b$, and the universal twist-2 collinear parton distribution functions (PDFs), $f_{a/A}$ (and $f_{b/B}$), to find a parton of flavor $a$ (and $b$) from the hadron $A$ (and $B$), as indicated in Eq.~(\ref{eq:fac-lp}).  The leading power contribution in Eq.~(\ref{eq:pexp}) contributes to the cross section, but, not to the SSA.  This is because the SSA is a naively time-reversal odd observable, we need,  in order to generate the SSA, a phase and a  spin-flip at the partonic scattering.   At the leading power in the collinear factorization, the phase could only be generated by the interference between the tree and the loop diagram, and the spin-flip could be achieved by the quark mass.  That is, the asymmetry generated in this way must be proportional to $\alpha_s m_Q/p_T$, which is a power suppressed small number, and is not sufficient to explain the data on the SSA\cite{Kane:1978nd}.   

In the QCD collinear factorization approach to SSAs, the asymmetry is generated by the quantum interference between the first and the second scattering amplitude in  Fig.~\ref{fig:xsec}, which is the first power correction to the spin-dependent cross section and is capable of generating the right phase and required spin flip between the single parton and the two-parton composite state to generate the SSA\cite{Efremov,qiu}.  It was argued in Ref.~[\refcite{QS-fac}] that such a power correction to the single jet/particle inclusive cross section of $A(p_A,s_\perp)+B(p_B) \to h(p)+X$ can be factorized in the same way as the leading power term, except that the PDF of the polarized hadron is replaced by a twist-3 quark-gluon correlation function $T_{(a,F)/A}(\vec{s})$ with $a=q,\bar{q}$, or $g$ as in Eq.~(\ref{eq:fac-Tf}), or a twist-2 transversity distribution $h_{a/A}(\vec{s})$ with $a=q$, or $\bar{q}$ to take care of the spin-flip while the other PDF is replaced by a spin-averaged twist-3 quark-gluon correlation function $T_{(b,F)/B}^{(\sigma)}$ to take care of the phase generation (or the twist-2 fragmentation function is replaced by corresponding twist-3 collinear fragmentation function\cite{Kang:2010zzb} in the case of inclusive single hadron production, which will not be discussed in this talk\cite{qiu}).  

The twist-3 distributions $T_{a,F}(x,x)$ and $T_{a,F}^{(\sigma)}(x,x)$ represent the long-distance effect of the quantum interference between a scattering amplitude with a single active parton and the one with an active two-parton composite state, as shown in Fig.~\ref{fig:xsec}.   Although the quantum interference between the partonic scattering amplitudes with different number of active partons is suppressed by a power of $1/Q$, its contribution to the SSA could be enhanced in certain parts of the phase space, in particular, in the forward region of the polarized hadron, which is a natural feature of twist-3 contributions\cite{qiu}.  The predictive power of the approach relies on the universality and our knowledge of the twist-3 quark-gluon correlation functions, which are defined in terms of matrix elements of twist-3 operators\cite{qiu,Kang:2008ey}, e.g., those in Eq.~(\ref{Tqs}).  Like the usual twist-2 PDFs, these twist-3 parton distributions relevant to the transverse SSAs are non-perturbative and universal, and should be extracted from experimental data on the SSAs. However, unlike the usual twist-2 PDFs, these distributions do not have the probability interpretation and are not necessarily positive.  On the other hand, we could interpret these distributions (or the matrix elements) as the expectation values of the inserted field operators  in Eq.~(\ref{Tqs}), which corresponds to the color Lorentz force and magnetic force experienced by the active parton\cite{Qiu:1993af}.

%%%%%%%%%%%%%%%%%%%%%%%%%%%%%%%%%
\section{Evolution and evolution kernels}

The evolution or the factorization scale dependence of these twist-3 parton correlation functions is an immediate consequence of the QCD factorization formalism for physical observables.   If one writes the spin-dependent cross section from the factorized formula in Eq.~(\ref{eq:fac-Tf}) or (\ref{eq:fac-Tfs}), schematically, as 
\begin{equation}
\Delta\sigma(Q,s_\perp) =
(1/Q) H^{\rm NLP} (Q/\mu,\alpha_s)\otimes f(\mu) \otimes {\cal T}(\mu) 
+ {\cal O}(1/Q^2)\, ,
\label{Dsigma_st}
\end{equation}
with the factorization scale $\mu$, one can derive from $d/d\ln(\mu)[ \Delta\sigma(Q,s_\perp) ]=0$ the leading order evolution equation for a generic twist-3 correlation function ${\cal T}$,
\begin{equation}
\frac{\partial}{\partial\ln(\mu)}{\cal T}
= \left(
\frac{\partial}{\partial\ln(\mu)} H^{{\rm NLP}(1)}-P_2^{(1)} \right)
\otimes {\cal T}\, ,
\label{eq:evolution}
\end{equation}
where $H^{{\rm NLP}(1)}$ and $P_2^{(1)}$ are the leading order partonic hard part and the evolution kernel of the twist-2 PDFs, respectively.  Both terms in the brackets on the right of Eq.~(\ref{eq:evolution}) are perturbatively calculable and give the leading order evolution kernel.   Since the distributions are universal, the perturbative evolution kernel could be derived in many different ways, although the results should be the same\cite{Kang:2008ey,Vogelsang:2009pj,Braun:2009mi}.  

\begin{figure}[h]
\begin{center}
$\frac{\partial}{\partial\ln \mu^2}$
\begin{minipage}[c]{2.9cm}
\includegraphics[width=2.9cm]{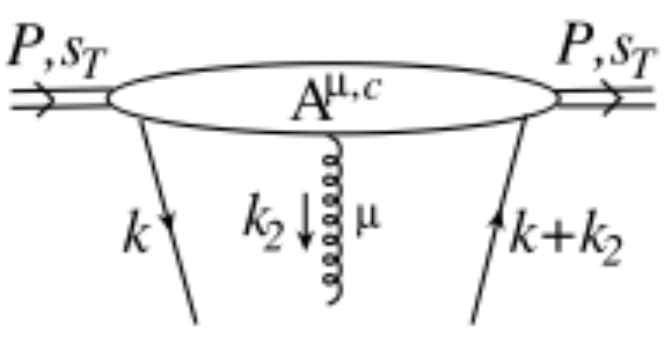}
\end{minipage}
$\approx $ \hskip 0.10in
\begin{minipage}[c]{2.2cm}
\includegraphics[width=2.2cm]{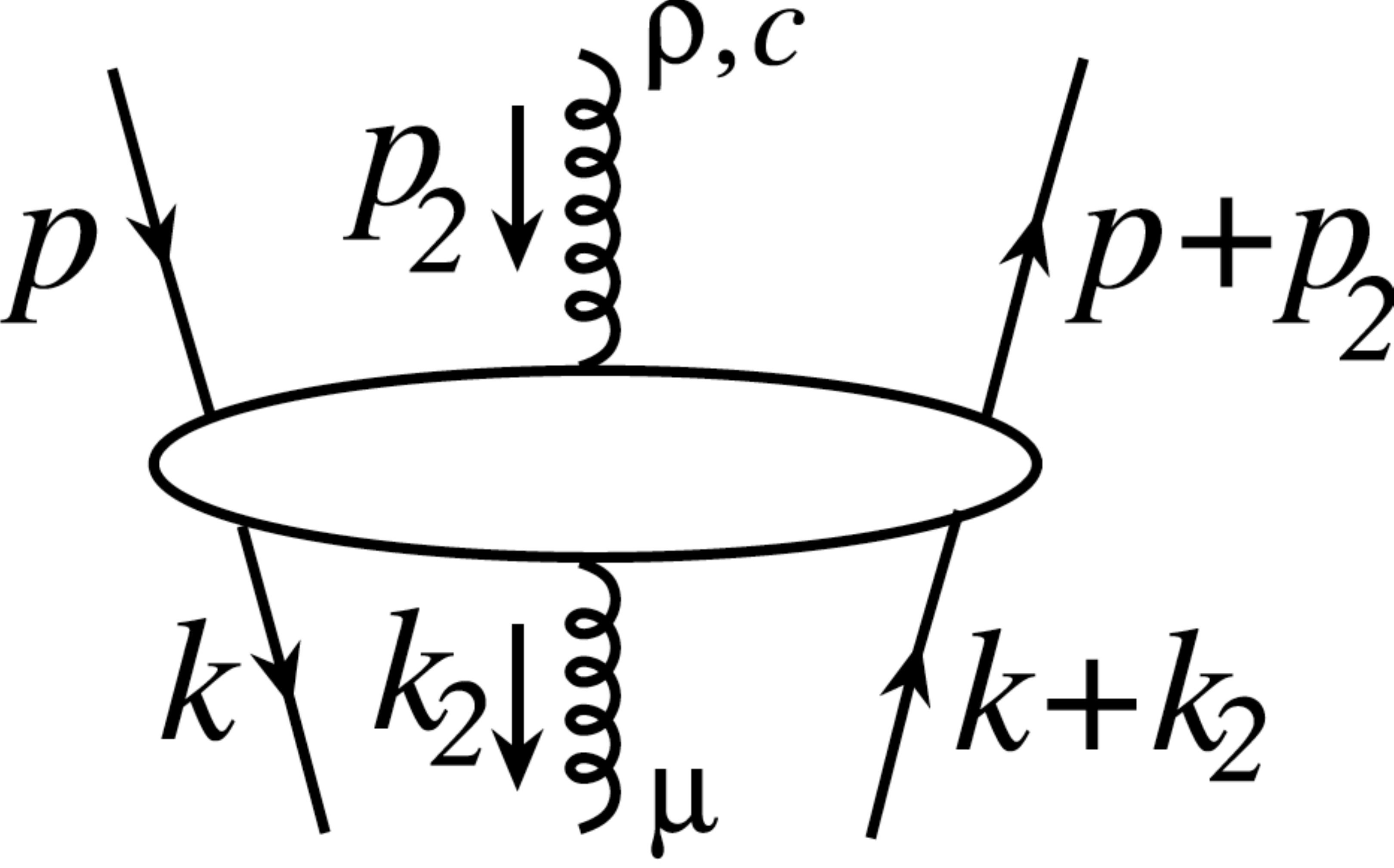}
\end{minipage}
$\ \otimes\ $ 
\begin{minipage}[c]{2.8cm}
\includegraphics[width=2.8cm]{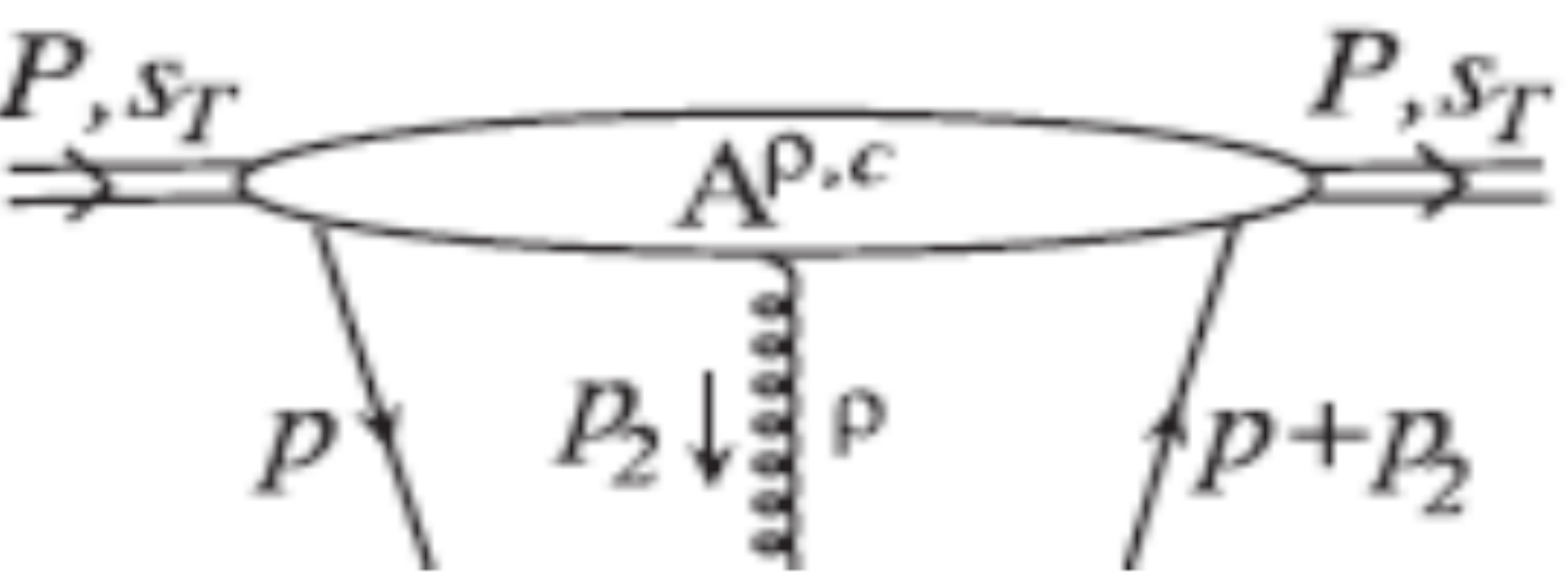}
\end{minipage}
\end{center}
%\vspace*{8pt}
\caption{A diagrammatic illustration for the evolution equation of the twist-3 quark-gluon correlation function. }
\label{fig:evo_q2q}
\end{figure}

In Ref.~[\refcite{Kang:2008ey}], we presented a derivation of the evolution equations of twist-3 quark-gluon correlation functions relevant to SSAs from the perturbative variation of these correlation functions.  The evolution equations can be schematically represented by the plot in Fig.~\ref{fig:evo_q2q}.  The evolution kernels can be perturbatively calculated from the first Feynman diagram on the right-hand-side of the diagrammatic equation in Fig.~\ref{fig:evo_q2q} with the bottom parton lines contracted by the cut vertices and the top parton lines contracted with the proper projection operators derived in Ref.~[\refcite{Kang:2008ey}].  For example, for the ETQS function $T_{q,F}(x_1,x_2)$, the cut vertex ${\cal V}_{q,F}$ and corresponding projection operator ${\cal P}_{q,F}$ in the light-cone (LC) gauge are given by 
\begin{eqnarray}
{\cal V}_{q,F}^{\rm (LC)}
&=&
\frac{\gamma^+}{2P^+}\,\delta\left(x-\frac{k^+}{P^+}\right)
x_2\,\delta\left(x_2-\frac{k_2^+}{P^+}\right)
\left(i\,\epsilon^{s_T\sigma n\bar{n}}\right)
\left[-g_{\sigma\mu}\right]\,
{\cal C}_q \, ,
\label{cv_q_lc}\\
{\cal P}_{q,F}^{(\rm LC)}
&=&
\frac{1}{2}\, \gamma\cdot P \left(\frac{-1}{\xi_2}\right) 
\left(i\, \epsilon^{s_T\rho\, n \bar{n}}\right) \, 
\widetilde{\cal C}_q\, ,
\label{proj_q_lc}  
\end{eqnarray}
where $\left({\cal C}_q\right)^c_{ij} = \left(t^c\right)_{ij}$ and $(\widetilde{\cal C}_q)^c_{ji}=2/(N_c^2-1) (t^c)_{ji}$ define the color contractions of quark and gluon fields, and the $\xi=p^+/P^+$ and $\xi_2=p_2^+/P^+$ are momentum fractions of the parent partons, as shown in Fig.~\ref{fig:evo_q2q}.

\begin{figure}[h]
\begin{center}
\includegraphics[width=0.98\columnwidth]{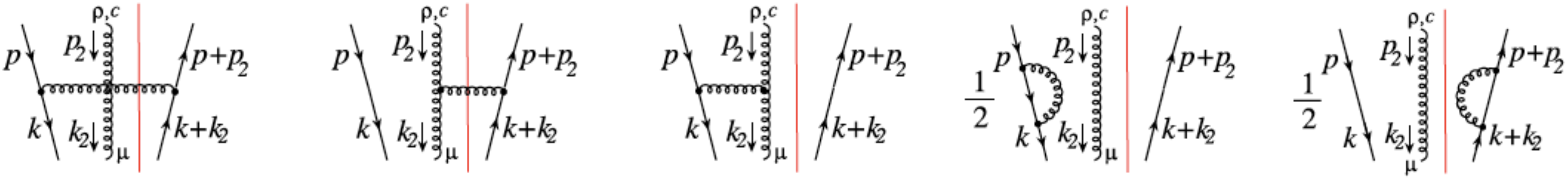}
\end{center}
\caption{Feynman diagrams contribute to the LO evolution kernel of quark-gluon correlation functions.}
\label{fig:loq}
\end{figure}

By calculating the leading order Feynman diagrams in Fig.~\ref{fig:loq} with the cut vertex and projection operator in Eqs. (9) and (10), respectively, we obtained finite contributions to the leading order evolution kernel and the following evolution equation\cite{Kang:2008ey},
\ben
\frac{\partial {T}_{q,F}(x,x,\mu)}{\partial{\ln \mu^2}}
&=&
\frac{\alpha_s}{2\pi}
\int_x^1\frac{d\xi}{\xi}
\bigg\{
P_{qq}(z)\, {T}_{q,F}(\xi,\xi,\mu)
\label{old}\\
&\ & \hskip -0.1in
+\frac{N_c}{2}
\left[\frac{(1+z){T}_{q,F}(\xi,x,\mu)
     -(1+z^2){T}_{q,F}(\xi,\xi,\mu)}{1-z}
     +{T}_{\Delta q,F}(x,\xi,\mu)\right]\bigg\},
\nonumber
\een
where $z=x/\xi$ and $P_{qq}(z)$ is the splitting kernel for unpolarized quark distribution function,
and the quark-gluon correlation function ${T}_{\Delta q,F}(x_1, x_2, \mu)$ is given by\cite{Kang:2008ey}
\ben
T_{\Delta q,F}(x_1, x_2)
&=&\int\frac{dy_1^- dy_2^-}{4\pi}\,
e^{i x_1 P^+y_1^-} e^{i (x_2-x_1) P^+y_2^-}
\label{TDqt}\\
&\ & \times
\langle P,s_T|\bar{\psi}_q(0)\,
\gamma^+\gamma^5
\left[i\, s_T^\alpha \, F_\alpha^{~ +}(y_2^-)\right] 
\psi_q(y_1^-)|P,s_T\rangle.
\nonumber
\een
The results derived in Refs.~[\refcite{Zhou:2008mz,Vogelsang:2009pj}] are consistent with ours.  But, the evolution equation derived later by Braun, Manashov, and Pirnay in Ref.~[\refcite{Braun:2009mi}] is slightly different,
\ben
\frac{\partial {T}_{q,F}(x,x,\mu)}{\partial{\ln \mu^2}}
&=&
\frac{\alpha_s}{2\pi}
\int_x^1\frac{d\xi}{\xi}
\bigg\{
P_{qq}(z)\, {T}_{q,F}(\xi,\xi,\mu)
\nnu
&\ &  \hskip -0.1in
+\frac{N_c}{2}
\left[\frac{(1+z){T}_{q,F}(\xi,x,\mu)
     -(1+z^2){T}_{q,F}(\xi,\xi,\mu)}{1-z}
     -{T}_{\Delta q,F}(x,\xi,\mu)\right]
\nnu
&\ &  %\hskip -0.1in
-N_c \, \delta(1-z)\, {T}_{q,F}(x, x, \mu)
\nonumber\\
&\ &
+\frac{1}{2N_c}\left[ (1-2 z) {T}_{q,F}(x, x-\xi, \mu) - {T}_{\Delta q,F}(x, x-\xi, \mu) \right]\bigg\}.
\label{braun}
\een

Comparing Eqs.~(\ref{old}) and (\ref{braun}), it is clear that two results differ by two contributions listed in the third and fourth line in Eq.~(\ref{braun}). In addition, there is a sign difference in front of the ${T}_{\Delta q,F}$ distribution in the second line. This sign difference is due to a fact that two groups used a different sign convention for anti-symmetric tensor $\epsilon^{\mu\nu\alpha\beta}$:  we chose $\epsilon^{0123}=1$, while Braun-Manashov-Pirnay used $\epsilon_{0123}=1$ implying $\epsilon^{0123}=-1$.  We also noticed that Ref.~[\refcite{Ma:2011nd}] used the same convention as that in our paper, thus they obtain the same sign for the ${T}_{\Delta q,F}$ term.

\begin{figure}[h]
\begin{center}
\includegraphics[width=1.0in]{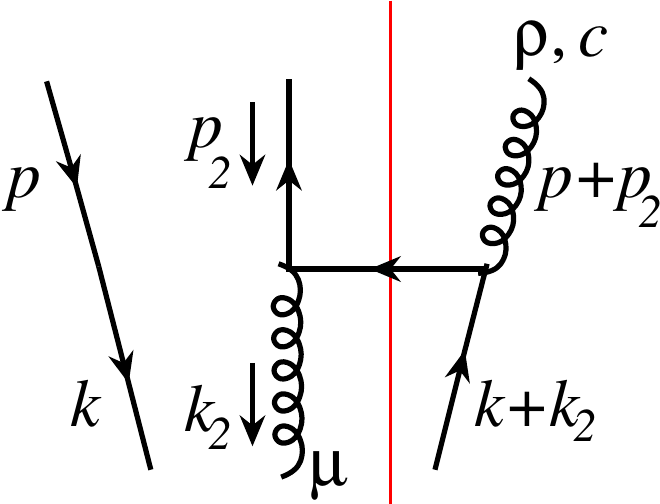}
\hskip 0.3in
\includegraphics[width=1.05in]{fig7r.pdf}
\caption{Feynman diagrams contribute to the evolution from the interference of a gluon and a quark-antiquark state (left), and the usual interference of a quark and a quark-gluon state (right).}
\label{extra}
\end{center}
\end{figure}
In the fourth line in Eq.~(\ref{braun}), the term $\propto 1/2N_c$  comes from the two Feynman diagrams in Fig.~\ref{extra}. The left diagram in Fig.~\ref{extra} corresponds to the interference between a gluon and a quark-antiquark pair. This diagram was missed in our original calculation\cite{Kang:2008ey}. After including this diagram, we obtain one half of the $1/2N_c$ term in Eq.~(\ref{braun}),
\ben
\left.\frac{\partial {T}_{q,F}(x,x,\mu)}{\partial{\ln \mu^2}}\right|_{\rm Fig.~\ref{extra}(left)} 
&=& \frac{\alpha_s}{2\pi}
\int_x^1\frac{d\xi}{\xi}
\frac{1}{2N_c}\left(\frac{1}{2}\right)\left[ (1-2 z) {T}_{q,F}(x, x-\xi, \mu) \right.
\nonumber\\
&\ & \hskip 1.3in \left.
+ {T}_{\Delta q,F}(x, x-\xi, \mu) \right],
\een
where again the ${T}_{\Delta q,F}$ term has an overall sign difference due to our convention for $\epsilon^{\mu\nu\alpha\beta}$. The right diagram in Fig.~\ref{extra} is actually Fig.~7(h) in our original paper Ref.~[\refcite{Kang:2008ey}]. This diagram vanishes if the quark on the left of the cut has a positive momentum $p^+=\xi P^+ > 0$, which was assumed in the original paper\cite{Kang:2008ey}. However, the $\xi$ does not have to be larger than 0 as long as  $\xi+\xi_2 > 0$.  By calculating the contribution from the region where $\xi < 0$, we find that it gives exactly the other half of the $1/2N_c$ term in Eq.~(\ref{braun}), and reproduce the fourth line of Eq.~(\ref{braun}) by adding contributions from both diagrams in Fig.~\ref{extra}.

The term in the third line in Eq.~(\ref{braun}), $-N_c T_{q,F}(x, x)$, was missed in our original paper\cite{Kang:2008ey}. The error was caused by a subtlety in taking the limit $x_2\to 0$ when we evaluate the integration $\int dx_2 \delta(x_2) x_2 F(x_2)=\lim_{x_2\to 0} x_2 F(x_2)$ to get the gluonic-pole matrix element.   The limit, $\lim_{x_2\to 0} x_2 F(x_2)$, would vanish if the function $F(x_2)$ is finite as $x_2\to 0$, which is unfortunately not always true in our calculation.  We find that Fig.~\ref{fig:loq}(b) and (c) have additional contributions to the evolution as\cite{Kang:2012em},
\ben
& \ & \frac{N_c}{2} \int_{x_2}^{1-x} 
d\xi_2 \,T_{q,F}(x, x+\xi_2, \mu) \left[- \frac{x_2}{\xi_2^2}\right] + \cdots, \ \mbox{and}
\label{7b}
\\
& \ & \frac{N_c}{2} \int_{x+x_2-1}^{x_2} 
d\xi_2 \,T_{q,F}(x+x_2-\xi_2, x+x_2, \mu) \left[\frac{x_2}{\xi_2^2}\right] + \cdots \, 
\label{7c}
\een 
respectively.
Here the ``$\cdots$'' includes any regular terms which vanish safely when we take $x_2\to 0$. The subtlety is caused by the fact that the integration over $d\xi_2$ in Eqs.~(\ref{7b}) and (\ref{7c}) is singular as $x_2\to 0$.  To evaluate the $d\xi_2$ integration, we first expand $T_{q,F}(x, x+\xi_2, \mu)$ around $\xi_2=x_2$ in Eq.~(\ref{7b}),
\ben
T_{q,F}(x, x+\xi_2) = T_{q,F}(x, x+x_2) + \frac{\partial}{\partial \xi_2}T_{q,F}(x, x+\xi_2)|_{\xi_2\to x_2} (\xi_2 - x_2) + \cdots \, .
\label{expansion}
\een
The integration in Eq.~(\ref{7b}) with the first term of the expansion in Eq.~(\ref{expansion}) gives 
\ben
&\ &
\frac{N_c}{2} \int_{x_2}^{1-x} 
d\xi_2 \,T_{q,F}(x, x+x_2, \mu) \left[- \frac{x_2}{\xi_2^2}\right] 
= \frac{N_c}{2}\,T_{q,F}(x, x+x_2, \mu)
\left[\frac{x_2}{\xi_2}\right]^{1-x}_{x_2} 
\nonumber \\
&\ & \hskip 0.5in
= \frac{N_c}{2}\,T_{q,F}(x, x+x_2, \mu)
\left[\frac{x_2}{1-x} - 1\right],
\een
which goes to $-\frac{N_c}{2} T_{q,F}(x, x, \mu)$ at the limit $x_2\to 0$. If one assumes $T_{q,F}(x, x+\xi_2, \mu)$ is a smooth (regular) function, we find that the higher order terms in the expansion in Eq.~(\ref{expansion}) do not contribute to the evolution in Eq.~(\ref{7b}) when $x_2\to 0$. So Fig.~\ref{fig:loq}(b) gives us an additional contribution $-\frac{N_c}{2} T_{q,F}(x, x, \mu)$. Similarly we find exactly the same contribution from Eq.~(\ref{7c}). Adding them together, we have
\ben
\left.\frac{\partial {T}_{q,F}(x,x,\mu)}{\partial{\ln \mu^2}}\right|_{\rm Fig.~\ref{fig:loq}(b+c)-additional} 
 = - N_c \,T_{q,F}(x, x, \mu),
\een
which reproduces the third line in Eq.~(\ref{braun}) and is what was missed in our original paper\cite{Kang:2008ey}. 

We now have a complete agreement with the Braun-Manashov-Pirnay result. Similarly, our results for flavor singlet evolution are also now consistent with the Braun-Manashov-Pirnay result.

%%%%%%%%%%%%%%%%%%%%%%%%%%%%%%%%%
\section{Evolution kernel of $T_{q,F}^{(\sigma)}$}

Using the same techniques, we could also derive the evolution equation for the other twist-three quark-gluon correlation function ${T}_{q,F}^{(\sigma)}(x,x,\mu)$. The calculation is straightforward, and the result is
\ben
\frac{\partial {T}_{q,F}^{(\sigma)}(x,x,\mu)}{\partial{\ln \mu^2}}
&=&
\frac{\alpha_s}{2\pi}
\int_x^1\frac{d\xi}{\xi}
\bigg\{
\Delta_{T}P_{qq}(z)\, {T}_{q,F}^{(\sigma)}(\xi,\xi,\mu)
\nonumber\\
&\ & \hskip 0.6in
+\frac{N_c}{2}
\left[\frac{2 \, {T}_{q,F}^{(\sigma)}(\xi,x,\mu)
     -2 z\, {T}_{q,F}^{(\sigma)}(\xi,\xi,\mu)}{1-z}\right]
\label{sigma}\\
&\ & \hskip 0.0in
-N_c \, \delta(1-z)\, {T}_{q,F}^{(\sigma)}(x, x, \mu)
+\frac{1}{2N_c} \left[2 (1-z) {T}_{q,F}^{(\sigma)}(x, x-\xi, \mu)\right] \bigg\},
\nonumber
\een
where $\Delta_{T}P_{qq}(z)$ is the splitting kernel for the quark transversity given by
\ben
\Delta_{T}P_{qq}(z) = C_F\left[\frac{2\, z}{(1-z)_+}+\frac{3}{2}\delta(1-z) \right].
\een
This evolution equation was first derived in Ref.~[\refcite{Zhou:2008mz}], which contains only the first two lines in Eq.~(\ref{sigma}). The first term in the third line, $-N_c\, {T}_{q,F}^{(\sigma)}(x, x, \mu)$, has exactly the same origin as those in Eqs.~(\ref{7b}) and (\ref{7c}) from calculating diagrams in Figs.~\ref{fig:loq}(b) and (c) with a caution of taking the limit $x_2\to 0$.  The second term, $\propto 1/2N_c$,  is again due to the fact that the Feynman diagrams in Fig.~\ref{extra} were not included in the calculation of Ref.~[\refcite{Zhou:2008mz}].

%%%%%%%%%%%%%%%%%%%%%%%%%%%%%%%%%
\section{Summary}

We have rederived the evolution equations for both ${T}_{q,F}(x, x, \mu)$ and ${T}_{q,F}^{(\sigma)}(x, x, \mu)$\cite{Kang:2012em}. We resolved the discrepancies in the literature for the evolution of ETQS function ${T}_{q,F}(x, x, \mu)$. We understand that such discrepancies were also resolved recently by the other two groups\cite{Schafer:2012ra,WV} through careful reexaminations of their original derivations in Refs.~[\refcite{Zhou:2008mz,Vogelsang:2009pj}], also in Ref.~[\refcite{Ma}] from a different approach. Using the same techniques developed in the current paper, we updated the calculation for the evolution of ${T}_{q,F}^{(\sigma)}(x, x, \mu)$ and found two additional contributions which are missing in the literature. These results will have important consequences, e.g., in the study of QCD resummation for the spin-dependent observables\cite{Kang:2011mr}.

%%%%%%%%%%%%%%%%%%%%%%%%%%%%%%%%%

%%%%%%%%%%%%%%%%%%%%%%%%%%%%%%%%%%%%%%%
\section*{Acknowledgments}

This work was supported in part by the US
Department of Energy, Office of Science, under Contract No. DE-AC52-06NA25396 and DE-AC02-98CH10886.

%%%%%%%%%%%%%%%%%%%%%%%%%%%%%%%%%%%%%%%
%\begin{thebibliography}{000} %for 3 digits
%\begin{thebibliography}{00}  %for 2 digits

%%%%%%%%%%%%%%%%%%%%%%%%%
\end{document}